\documentclass[aps,showpacs]{revtex4}
\newcommand{\be}{\begin{equation}}
\newcommand{\ee}{\end{equation}}
\newcommand{\ben}{\begin{eqnarray}}
\newcommand{\een}{\end{eqnarray}}
\newcommand{\bb}{\bibitem}
\newcommand{\ov}{\overline}
\newcommand{\wt}{\widetilde}
\begin{document}

\title{Lorentz and CPT symmetries in commutative and noncommutative
spacetime}

\author{D. Bazeia, T. Mariz, J.R. Nascimento, E. Passos, and R.F. Ribeiro} 

\affiliation{Departamento de F\'\i sica, Universidade Federal da
Para\'\i ba, Caixa Postal 5008, 58051-970 Jo\~ao Pessoa, PB, Brazil}

\date{\today}

\begin{abstract}
We investigate the fermionic sector of a given theory, in which massive and
charged Dirac fermions interact with an Abelian gauge field, including a
non standard contribution that violates both Lorentz and CPT symmetries.
We offer an explicit calculation in which the radiative corrections due to
the fermions seem to generate a Chern-Simons-like effective action. Our
results are obtained under the general guidance of dimensional regularization,
and they show that there is no room for Lorentz and CPT violation in both
commutative and noncommutative spacetime.
\end{abstract}

\pacs{11.30.Cp, 11.30.Er}

\maketitle

Maxwell's theory of electromagnetism was crucial to question
Galileu invariance, to give rise to Lorentz symmetry. Nowadays, in string
theory one may find a way to question Lorentz invariance, since there are
interactions that support spontaneous breaking of Lorentz symmetry \cite{ksa}.
In string theory, one may also find room for noncommutativity of the
coordinates that define the spacetime manifold \cite{sw}. Thus, is appears
legitimate to investigate possible breaking of Lorentz invariance
in both commutative and noncommutative spacetime.

The issue of breaking Lorentz invariance has been recently addressed by
many authors. The standard route \cite{cfj,gt,cf,ck} includes a modification
of Maxwell's theory, in which one adds the Chern-Simons-like term
$\kappa_{\mu}\varepsilon^{\mu\nu\lambda\rho}F_{\nu\lambda}A_{\rho}$.
The problem relies on recognizing that Lorentz and CPT symmetries are violated
in the fermionic sector of a given theory, which contains the contributions
\cite{1,2,3,4,5,6,7,8,9,10,11,12,13}
\be\label{faction}
I_f=\int d^4x\;{\overline\psi}\left(i/\!\!\!\partial-m-
{/\!\!\!\!A}-{/\!\!\!b}\gamma_5\right)\psi
\ee
The three first terms are usual; they describe charged
and massive Dirac fermions coupled to an Abelian gauge field. However,
the fourth term is unusual: $b_{\mu}$ is a constant four vector which
selects a fixed direction in spacetime, and explicitly
violates Lorentz and CPT symmetries. The fermions can be integrated,
and the radiative result may lead to
\be\label{cs}
I_{CS}=\frac12\int\,d^4x\,\varepsilon^{\mu\nu\lambda\rho}\,\kappa_{\mu}
F_{\nu\lambda}A_{\rho}
\ee
with $\kappa_{\mu}$ being proportional to $b_{\mu}$, that is
$\kappa_{\mu}=C\,b_{\mu}$. This result, if correct,
introduces a modification of electrodynamics, which allows for the explicit
violation of Lorentz and CPT symmetries. The issue has been carefully
investigated in several different contexts, leading to results in which $C$
vanishes \cite{1,8} or not \cite{2,3,4,5,6,7,9,10,11,12,13}.

In the present work we revisit the problem, with the aim to extend the
calculation to the noncommutative spacetime manifold. The importance of
investigating noncommutativity of spacetime has been brought to high energy
physics via string theory \cite{sw} -- see also \cite{filk,jac} for other
informations. In our quest to deal with the issue in the standard
case, however, we had to introduce new calculations which led us to the result
that there is neither Lorentz nor CPT violation in the commutative spacetime.
And this was also shown to be correct in the noncommutative case. These results
were obtained under the general guidance of dimensional regularization, and
they have led us to offer our calculations in a form as standard as possible,
keeping track of the main steps and enlightening the way the puzzle shows up:
as we shall see, there is an intricate entanglement between the calculation
involving the Dirac matrices and the evaluation of the momentum integral
of all the contributions at first order in the vector that responds for
Lorentz and CPT violation, and at second and third (in the noncommutative
case) order in the gauge field. We implement our investigations using
derivative expansion of operators \cite{de1,de2,de3,de4,de5}, and we consider
the spacetime as commutative and noncommutative.

Firstly we work in the commutative case. To account for the fermionic
integration we write
\be
e^{i I[b,A]}=\int D{\ov \psi}D\psi\, e^{i\int d^4x{\cal L}_f}
\ee
where the effective action is given by
\be
I[b,A]=-i\,{\rm Tr}\,\ln({/\!\!\!p}-m-{/\!\!\!\!A}-{/\!\!\!b}\gamma_5)
\ee
We use this expression to write: $I[b,A]=I[b]+I^\prime[b,A]$. The first
term is $I[b]=-i\,{\rm Tr}\ln({/\!\!\!p}-
m-{/\!\!\!b}\gamma_5)$, which does not depend on the gauge field.
The second term is $I^{\,\prime}[b,A]$, which is given by
\be\label{ea}
I^{\,\prime}[b,A]=i\,{\rm Tr} \sum_{n=1}^{\infty}\frac1n
\Biggl[\frac1{{/\!\!\!p}-m-{/\!\!\!b}\gamma_5}{/\!\!\!\!A}\Biggr]^n
\ee
In this expression we single out the term
\be
I^{(2)}[b,A]=\frac{i}{2}{\rm Tr}\frac{1}{/\!\!\!p-m-/\!\!\!b
\gamma_5}\;{/\!\!\!\!A}\;\frac{1}{/\!\!\!p-m-/\!\!\!b\gamma_5}\;{/\!\!\!\!A}
\ee
which is the term that matter, in the quest to find how the radiative
corrections generate the Chern-Simons-like term written in Eq.~(\ref{cs}).

We can proceed following two distinct routes, in which one includes or not
the contribution involving the vector $b_{\mu}$ into the Dirac
propagator -- see Ref.~{\cite{2}} for details. In the present work we
follow the perturbative route, so we use the expression
\be\label{exp}
\frac{1}{/\!\!\!p-m-/\!\!\!b\gamma_5}=\frac{1}{/\!\!\!p-m}+\frac{1}
{/\!\!\!p-m}\;{/\!\!\!b\gamma_5}\;\frac{1}{/\!\!\!p-m}+\cdots
\ee
to write, to first order in $b$ and second order in $A$
\be\label{eat}
I^{(1,2)}[b,A]=\frac{i}{2}\;{\rm Tr}
\bigl[S(p)\;{/\!\!\!b\gamma_5}\;S(p)\;
{/\!\!\!\!A}\;S(p)\;{/\!\!\!\!A}+S(p)\;{/\!\!\!\!A}\;S(p)\;
{/\!\!\!b\gamma_5}\;S(p)\;{/\!\!\!\!A}\bigr]
\ee
where we have set
\be
S(p)=\frac1{/\!\!\!p-m}
\ee
We rewrite Eq.~(\ref{eat}) in the form
\be\label{ea1}
I^{(1,2)}[b,A]=\frac{i}{2}\int d^4x\;\left(\Pi_a^{\mu\nu}+
\Pi_b^{\mu\nu}\right)\,A_{\mu}A_{\nu}
\ee
where
\be
\Pi_a^{\mu\nu}={\rm tr}\int\frac{d^4p}{(2\pi)^4} S(p)\;{/\!\!\!b}
\gamma_5\;S(p)\;\gamma^{\mu}\;S(p-i\partial)\;\gamma^{\nu}
\ee
and
\be
\Pi_b^{\mu\nu}={\rm tr}\int\frac{d^4p}{(2\pi)^4} S(p)\;\gamma^{\mu}\;
S(p-i\partial) \;{/\!\!\!b}\gamma_5\;S(p-i\partial)\;\gamma^{\nu}
\ee
where ${\rm tr}$ stands for the trace over the Dirac matrices.

We now follow Refs.~{\cite{de1,de2,de3,de4,de5} and use the expansion
\be\label{iden}
\frac1{{/\!\!\!p}-i{/\!\!\!\partial}-m}=\frac1{{/\!\!\!p}-m}+
\frac1{{/\!\!\!p}-m}\;i{/\!\!\!\partial}\;\frac1{{/\!\!\!p}-m}+\cdots
\ee
which is valid up to first order in $\partial$, which is the expression
we need to generate the Chern-Simons-like term. With this we
change $\Pi_a^{\mu\nu}\to\Pi_1^{\mu\nu}$ and rewrite it in the form
\be\label{pi1}
\Pi_{1}^{\mu\nu}={\rm tr}\int\frac{d^4p}{(2\pi)^4}\;S(p){/\!\!\!b}\gamma_5S(p)
\gamma^{\mu}S(p)i{/\!\!\!\partial}S(p)\gamma^{\nu}
\ee
Also, we change $\Pi_b^{\mu\nu}\to\Pi_2^{\mu\nu}+\Pi_3^{\mu\nu}$ to write
\be\label{pi21}
\Pi_2^{\mu\nu}={\rm tr}\int\frac{d^4p}{(2\pi)^4}\;S(p)
\gamma^{\mu}S(p){/\!\!\!b}\gamma_5S(p)i{/\!\!\!\partial}S(p)\gamma^{\nu}
\ee
and
\be\label{pi22}
\Pi_3^{\mu\nu}={\rm tr}\int\frac{d^4p}{(2\pi)^4}\;S(p)\gamma^{\mu}
S(p)i{/\!\!\!\partial}S(p){/\!\!\!b}\gamma_5S(p)\gamma^{\nu}
\ee

We work with $\Pi_1^{\mu\nu}$. It can be written as
$\Pi_1^{\mu\nu}=\Pi_{1,{\rm div}}^{\mu\nu}+\Pi_{1,{\rm fin}}^{\mu\nu}$, where
\be
\Pi_{1,{\rm div}}^{\mu\nu}=i\,b_{\lambda}\,{\rm tr}\int\frac{d^4p}{(2\pi)^4}\;
\frac{{/\!\!\!p}\gamma^{\lambda}\gamma_5{/\!\!\!p}\gamma^{\mu}
{/\!\!\!p}{/\!\!\!\partial}{/\!\!\!p}\gamma^{\nu}}{(p^2-m^2)^4}
\ee
and
\ben\label{pi1fin}
\Pi_{1,{\rm fin}}^{\mu\nu}&=&i\,m^2\,b_{\lambda}\,{\rm tr}\int
\frac{d^4p}{(2\pi)^4}\;\frac{1}{(p^2-m^2)^4}(\,{/\!\!\!p}\gamma^{\lambda}
\gamma_5{/\!\!\!p}\gamma^{\mu}{/\!\!\!\partial}\gamma^{\nu}+{/\!\!\!p}
\gamma^{\lambda}\gamma_5\gamma^{\mu}{/\!\!\!p}{/\!\!\!\partial}
\gamma^{\nu}+\nonumber
\\
&&{/\!\!\!p}\gamma^{\lambda}\gamma_5\gamma^{\mu}{/\!\!\!\partial}
{/\!\!\!p}\gamma^{\nu}+\gamma^{\lambda}\gamma_5{/\!\!\!p}\gamma^{\mu}
{/\!\!\!p}{/\!\!\!\partial}\gamma^{\nu}+\gamma^{\lambda}\gamma_5
{/\!\!\!p}\gamma^{\mu}{/\!\!\!\partial}{/\!\!\!p}\gamma^{\nu}+
\gamma^{\lambda}\gamma_5\gamma^{\mu}{/\!\!\!p}{/\!\!\!\partial}{/\!\!\!p}
\gamma^{\nu}+m^2\;{\gamma^{\lambda}\gamma_5\gamma^{\mu}{/\!\!\!\partial}
\gamma^{\nu}})
\een
The other two terms $\Pi_2^{\mu\nu}$ and $\Pi_3^{\mu\nu}$ are similar,
and are treated similarly.

We evaluate the integrals under the general guidance of dimensional
regularization \cite{dr1,dr2,dr3}. Thus, we change dimensions from
$4$ to $2w$, and we change $d^4p/(2\pi)^4$ to
$(\mu^2)^{2-w}[d^{2w}p/(2\pi)^{2w}]$, where $\mu$ is an arbitrary
parameter that identifies the mass scale. We use two distinct routes
to do the calculations involving the Dirac matrices. In the first route
we use the ciclic property of the trace, to move $\gamma_5$ to the very
end of every expression involving the trace of Dirac matrices. The
potential divergences in the momentum integration come from the first
term of $\Pi_{1}^{\mu\nu}$. We use
\be
\int\frac{d^{2w}p}{(2\pi)^{2w}}\;\frac{p_{\alpha}\,p_{\beta}\,p_{\gamma}\,
p_{\delta}}{(p^2-m^2)^4}=\frac{i}{24(4\pi)^w}
\;\frac{\Gamma(2-w)}{(m^2)^{2-w}}G_{\alpha\beta\gamma\delta}
\ee
where $G_{\alpha\beta\gamma\delta}=g_{\alpha\beta}g_{\gamma\delta}+
g_{\alpha\gamma}g_{\beta\delta}+g_{\alpha\delta}g_{\beta\gamma}$.
We also use $\{\gamma^{\alpha},\gamma^{\beta}\}=2g^{\alpha\beta}$ and
$\gamma^{\alpha}\gamma_{\alpha}=2w$ in order to
rewrite Eq.~(\ref{ea1}) in the form
\be\label{ifinal}
I^{(1,2)}[b,A]=\frac32\,i\,\Pi(w)\,b_{\mu}\,{\rm tr}(\gamma^{\mu}
\gamma^{\nu}\gamma^{\lambda}\gamma^{\rho}\gamma_5)\int d^4x\,
\partial_{\nu}A_{\lambda}A_{\rho}
\ee
Here the factor $3$ accounts for identical contributions
that comes from $\Pi_1^{\mu\nu}$, $\Pi_2^{\mu\nu}$ and
$\Pi_3^{\mu\nu}$. Also, $\Pi(w)$ is given by
\be\label{piw}
\Pi(w)=-\frac{2w-1}{96\pi^2}+\frac{w+1}{96\pi^2}\left(\frac{4\pi\mu^2}{m^2}
\right)^{2-w}\Gamma(2-w)(2-w)
\ee

In the above calculations we have set
$\Pi^{\mu\nu}_1=\Pi^{\mu\nu}_{1,{\rm div}}+\Pi^{\mu\nu}_{1,{\rm fin}}$
to split the $\Pi^{\mu\nu}_1$ contribution into two parts, one divergent
and the other finite. The contribution $\Pi_{1,{\rm div}}^{\mu\nu}$
is divergent in the limit $w\to2$, and it contribute with the term
proportional to $\Gamma(2-w)$. However, the factor involving the Dirac
matrices contributes with the term $(2-w)$, in a way such that the full
contribution is finite in the limit $w\to2$. Furthermore, this finite term
exactly compensates the finite contribution that appears from
$\Pi_{1,{\rm fin}}^{\mu\nu}$ in the limit $w\to2$. In the limit
$w\to2$ we can use 
${\rm tr}(\gamma^{\mu}\gamma^{\nu}\gamma^{\lambda}\gamma^{\rho}\gamma_5)
=4i\varepsilon^{\mu\nu\lambda\rho}$, but $\Pi(w\to2)\to0$ and this
leaves no room for Lorentz and CPT violation. The perfect balance between
the two contributions that we have just found has been identified before
in Ref.~{\cite{cgs}} as being peculiar to dimensional regularization. We
stress that if one uses the relation $\{\gamma^\mu,\gamma_5\}=0$ to move
$\gamma_5$ to the end of every expression involving the trace of Dirac
matrices, the perfect balance between the two contributions is broken,
giving rise to a non zero value for the constant $C$. This result is
due to the use of $\{\gamma^\mu,\gamma_5\}=0$, which is valid in the
four dimensional spacetime, but we are working in $2w$ dimensions.

We make this point stronger by considering another route to implement the
calculation involving properties of the Dirac matrices when the spacetime
has dimension $2w$. We follow \cite{dr3,bm,bu,msr} and now the Dirac matrices
corresponding to the external indices $\mu,\nu$ and $\lambda$ are physical
matrices; they are written in the form ${\bar\gamma}^{\mu}$, etc.
The contribution
\be
{\rm tr}(\gamma^{\alpha}{\bar\gamma}^{\lambda}\gamma_5\gamma^{\beta}
{\bar\gamma}^{\mu}\gamma^{\gamma}\gamma^{\rho}\gamma^{\delta}
{\bar\gamma}^{\nu})G_{\alpha\beta\gamma\delta}
\ee
splits into the three terms
\be
{\rm tr}(\gamma^{\alpha}{\bar\gamma}^{\lambda}\gamma_5\gamma_{\alpha}
{\bar\gamma}^{\mu}\gamma^{\beta}\gamma^{\rho}\gamma_{\beta}{\bar\gamma}^{\nu})
+{\rm tr}(\gamma^{\alpha}{\bar\gamma}^{\lambda}\gamma_5\gamma^{\beta}
{\bar\gamma}^{\mu}\gamma_{\alpha}\gamma^{\rho}\gamma_{\beta}{\bar\gamma}^{\nu})
+{\rm tr}(\gamma^{\alpha}{\bar\gamma}^{\lambda}\gamma_5\gamma^{\beta}
{\bar\gamma}^{\mu}\gamma_{\beta}\gamma^{\rho}\gamma_{\alpha}{\bar\gamma}^{\nu})
\ee
and the Dirac matrices are changed according to
$\gamma^{\alpha}\to{\bar\gamma}^{\alpha}+{\hat\gamma}^{\alpha}$, where
$\{{\bar\gamma}^{\alpha},{\bar\gamma}^{\beta}\}=2{\bar g}^{\alpha\beta}$,
$\{{\hat\gamma}^{\alpha},{\hat\gamma}^{\beta}\}=2{\hat g}^{\alpha\beta}$,
and $\{{\bar\gamma}^{\alpha},{\hat\gamma}^{\beta}\}=0$, and also
${\bar\gamma}^{\alpha}{\bar\gamma}_{\alpha}=4$,
${\bar\gamma}^{\alpha}{\hat\gamma}_{\alpha}=0$ and
${\hat\gamma}^{\alpha}{\hat\gamma}_{\alpha}=2(w-2)$.
In this case we can use either the ciclic property of the trace, or the
relations $\{\gamma_5,{\bar\gamma}^{\mu}\}=[\gamma_5,{\hat\gamma}^{\mu}]=0$
to rewrite Eq.~(\ref{ea1}) in the form
\be
I^{(1,2)}[b,A]=\frac32\,\Pi^{\prime}(w)\,b_{\mu}\,\int d^4x\,
\varepsilon^{\mu\nu\lambda\rho}\,\partial_{\nu}A_{\lambda}A_{\rho}
\ee
where we have used
${\rm tr}({\bar\gamma}^{\mu}{\bar\gamma}^{\nu}
{\bar\gamma}^{\lambda}{\bar\gamma}^{\rho}\gamma_5)=4i
\varepsilon^{\mu\nu\lambda\rho}$ and ${\rm tr}({\gamma}^{\mu}{\gamma}^{\nu}
{\gamma}^{\lambda}{\hat\gamma}^{\rho}\gamma_5)=0$. Also,
$\Pi^{\prime}(w)=-4\Pi(w)$. We use this result to
write Eq.~(\ref{ifinal}) as in the Chern-Simons-like term in Eq.~(\ref{cs}),
where $\kappa_{\mu}=C\,b_{\mu}$, with $C$ given by
\be\label{C}
C=\frac{2w-1}{16\pi^2}-\frac{1+w}{16\pi^2}
\left(\frac{4\pi\mu^2}{m^2}\right)^{2-w}\Gamma(2-w)(2-w)
\ee
We see that $C\to0$ in the limit $w\to2$, which confirms the former result,
in which we have used the ciclic property of the trace.

We now consider the spacetime noncommutative
\cite{nc1,nc2,nc3,nc4,nc5,nc6,nc7,nc8}. In this case we set
$[x^{\mu},x^{\nu}]=i\theta^{\mu\nu}$, where $\theta^{\mu\nu}$ is a constant
antisymmetric tensor. As a consequence, one replaces the ordinary product of
functions by the Moyal product
\be
f(x)\star g(x)=e^{\frac{i}{2}\theta^{\mu\nu}\partial_{\mu}
\partial^{\prime}_{\nu}}\;
f(x)g(x^{\prime})\Bigl|_{x^{\prime}=x}
\ee
The first modification we have
to introduce concerns Eq.~(\ref{faction}), which should be changed to
\be\label{fnc}
{\wt I}_f=\int d^4x {\bar\psi}\star(i{/\!\!\!\partial}-m-{/\!\!\!b}
\gamma_5-{/\!\!\!\!A}\star)\psi
\ee
or better
\be
{\wt I}_f=\int d^4x {\bar\psi}(x)(i{/\!\!\!\partial}^{\,\prime}-m-{/\!\!\!b}
\gamma_5-e^{i\partial\times\partial^{\prime}}{/\!\!\!\!A})\psi(x^{\prime})
\Big|_{x^{\prime}=x}
\ee
where we are working with the fundamental (or anti-fundamental) representation,
of the gauge group, using $(1/2)\theta^{\mu\nu}\partial_{\mu}
\partial_{\nu}^{\prime}=\partial\times\partial^{\prime}$. In this
case Eq.~(\ref{cs}) should be changed to
\be\label{csnc}
{\wt I}_{CS}=\int\,d^4x\,\varepsilon^{\mu\nu\lambda\rho}
{\wt\kappa}_{\mu}\left({\partial}_{\nu}A_{\lambda}\star A_{\rho}+\frac{2}{3}i\,
A_{\nu}\star A_{\lambda}\star A_{\rho}\right)
\ee
where $\wt\kappa_{\mu}$ must have the form $\wt\kappa_{\mu}={\wt C}\,b_{\mu}$,
to include modifications coming from the noncommutativity of spacetime.

We are working with the fundamental representation of the gauge group.
In this case noncommutativity seems to appear changing the gauge field
${A}$ by the new field ${\wt A}=e^{\partial\times p}\,A$, and this
identification ease the work we have to implement, since we now see
that the above modification changes Eq.~(\ref{ea}) to
\be\label{I}
{\wt I}^{\prime}[b,A]=i\,{\rm Tr} \sum_{n=1}^{\infty}\frac1n
\Biggl[\frac1{{/\!\!\!p}-m-{/\!\!\!b}\gamma_5}
e^{\partial\times p}{/\!\!\!\!A(x)}\Biggr]^n
\ee
and now we single out the term
\be
{\wt I}^{(2)}[b,A]=\frac{i}2\,{\rm Tr}\Biggl[\frac1{{/\!\!\!p}-m-{/\!\!\!b}
\gamma_5}e^{\partial\times p}{/\!\!\!\!A(x)}\frac1{{/\!\!\!p}-m-{/\!\!\!b}
\gamma_5}e^{\partial^{\prime}\times p}
{/\!\!\!\!A(x^{\prime})}\Biggr]_{x^{\prime}=x}
\ee
which modifies the former calculations as follows: we rewrite
Eq.~(\ref{ea1}) in the form
\be\label{ea2}
{\wt I}^{(1,2)}[b,A]=\frac{i}2 \int d^4x \left({\wt\Pi}_a^{\mu\nu}+
{\wt\Pi}_b^{\mu\nu}\right) A_{\mu}\star A_{\nu}^{\prime}\Bigl|_{x^{\prime}=x}
\ee
where the terms ${\wt\Pi}_a^{\mu\nu}$ and ${\wt\Pi}_b^{\mu\nu}$
are now given by
\be
{\wt\Pi}_a^{\mu\nu}={\rm tr}\int\frac{d^4p}{(2\pi)^4} S(p)\;{/\!\!\!b}
\gamma_5\;S(p)\gamma^{\mu}S(p-i\partial)\gamma^{\nu}
e^{(\partial+\partial^{\prime})\times p}
\ee
and
\be
{\wt\Pi}_b^{\mu\nu}={\rm tr}\int\frac{d^4p}{(2\pi)^4} S(p)\gamma^{\mu}
S(p-i\partial)\;{/\!\!\!b}\gamma_5S(p-i\partial)\gamma^{\nu}
e^{(\partial+\partial^{\prime})\times p}
\ee
In both cases, expanding the phase factors and propagators up to first order
in the derivative the result adds to zero, like it did in the commutative case.
Thus, in the Chern-Simons-like contribution that appear in Eq.~(\ref{csnc}),
the term proportional to $\partial_{\mu}A_{\nu}\star A_{\lambda}$ remains
as in the former result in the commutative case.

In the noncommutative case there is another contribution, trilinear
in the gauge field, which comes from Eq.~(\ref{I}) for $n=3$. This
contribution is given by
\be
{\wt I}^{(3)}[b,A]=\frac{i}{3}{\rm Tr}\Biggl[\frac{1}
{/\!\!\!p-m-{/\!\!\!b}\gamma_5}\,e^{\partial\times p}\,{/\!\!\!\!A}(x)
\frac{1}{/\!\!\!p-m-{/\!\!\!b}\gamma_5}\,e^{\partial^{\prime}\times p}
\,{/\!\!\!\!A}(x^{\prime})\frac{1}{/\!\!\!p-m-{/\!\!\!b}\gamma_5}
\,e^{\partial^{\prime\prime}\times p}\,{/\!\!\!\!A}(x^{\prime\prime})
\Biggr]_{x^{\prime\prime}=x^{\prime}=x}
\ee
We use Eq.~(\ref{exp}) to write, selecting the terms that are linear
in $b$,
\ben
{\wt I}^{(1,3)}[b,A]&=&\frac{i}3 {\rm Tr}S(p)\Bigl[e^{{\partial}\times p}
{/\!\!\!\!A}(x)S(p)e^{\partial^{\prime}\times p}{/\!\!\!\!A}(x^{\prime})S(p)
{/\!\!\!b}\gamma_5S(p)+e^{{\partial}\times p}
{/\!\!\!\!A}(x)S(p){/\!\!\!b}\gamma_5S(p)e^{{\partial}^{\prime}\times p}
{/\!\!\!\!A}(x^{\prime})S(p)+\nonumber
\\
&&{/\!\!\!b}\gamma_5S(p)e^{{\partial}\times p}
{/\!\!\!\!A}(x)S(p)e^{{\partial}^{\prime}\times p}
{/\!\!\!\!A}(x^{\prime})S(p)\Bigr]
e^{{\partial}^{\prime\prime}\times p}
{/\!\!\!\!A}(x^{\prime\prime})\Bigl|_{x^{\prime\prime}=x^{\prime}=x}
\een
We use the identity (\ref{iden}) to write
\be\label{i31}
{\wt I}^{(1,3)}[b,A]=\frac{i}{3}\int d^4x \sum_{n=1}^3
\Gamma_n^{\,\mu\rho\nu}e^{(\partial+\partial^{\prime}+\partial^{\prime\prime})
\times p}A_{\mu}\star A^{\prime}_{\rho}\star A^{\prime\prime}_{\nu}
\Bigl|_{x^{\prime\prime}=x^{\prime}=x}
\ee
where
\be\label{G1}
\Gamma_1^{\,\mu\rho\nu}={\rm tr}\int\frac{d^4p}{(2\pi)^4}\;
S(p){/\!\!\!b}\gamma_5S(p)\gamma^{\mu}S(p)\gamma^{\rho}S(p)\gamma^{\nu}
\ee
and
\be\label{G2}
\Gamma_2^{\,\mu\rho\nu}={\rm tr}\int\frac{d^4p}{(2\pi)^4}\;
S(p)\gamma^{\mu}S(p){/\!\!\!b}\gamma_5S(p)\gamma^{\rho}S(p)\gamma^{\nu}
\ee
and
\be\label{G3}
\Gamma_3^{\,\mu\rho\nu}={\rm tr}\int\frac{d^4p}{(2\pi)^4}\;
S(p)\gamma^{\mu}S(p)\gamma^{\rho}S(p){/\!\!\!b}\gamma_5S(p)\gamma^{\nu}
\ee

These three terms are very similar to the three terms $\Pi_1^{\mu\nu}$,
$\Pi_2^{\mu\nu}$ and $\Pi_3^{\mu\nu}$ that we have found in
Eqs.~(\ref{pi1}), (\ref{pi21}) and (\ref{pi22}) of the former calculation.
They contribute similarly, and we can write, expanding the phase factor
up to zeroth order in the derivative,
\be
{\wt I}^{(1,3)}[b,A]=i\Gamma(w)\int d^{4}x\,b_{\mu}\,
\varepsilon^{\mu\nu\lambda\rho}\,A_{\nu}\star A_{\lambda}\star A_{\rho}
\ee
where $\Gamma(w)=-4\Pi(w)$ -- see Eq.~(\ref{piw}).
We then add ${\wt I}^{(1,2)}[b,A]$ and ${\wt I}^{(1,3)}[b,A]$ to write
Eq.~(\ref{csnc}) in the form
\be
{\wt I}_{CS}=C\,\int\,d^{4}x\,\varepsilon^{\mu\nu\lambda\rho}
b_{\mu}\left(\partial_{\nu}A_{\lambda}\star A_{\rho}+\frac23\,i\,
A_{\nu}\star A_{\lambda}\star A_{\rho}\right)
\ee
where $C$ is given in Eq.~(\ref{C}), the same result obtained in the
commutative case. Thus, in the limit $w\to2$ there is no room
for Lorentz and CPT violation also in the noncommutative case that we
have just considered.

We can also work with the adjoint representation of the gauge group. In this
case, in the fermionic action in Eq.~(\ref{ea}) we should change $A$ to
${\tilde A}_{ad}=(e^{\partial\times p}-e^{-\partial\times p})A$; also, we
include an extra factor of $1/2$ in this fermionic action, in order to
account for the use of Majorana spinors. The change in the gauge field
is similar to the identification done in the former case,
for the fundamental representation. The calculation is similar, and the
procedure that we follow is: the phase factors that appear from non-planar
diagrams are also expanded up to first order in the derivative, in order
to maintain the original program of searching for contributions linear
in the derivative and bilinear in the gauge field, and trilinear in the
gauge field. Within this context, the result in Eq.~(\ref{ea2}) should
be changed to
\be\label{ea21}
{\wt I}^{(1,2)}_{ad}[b,A]=\frac{i}2 \int d^4x \left({\wt\Pi}_{a,ad}^{\mu\nu}+
{\wt\Pi}_{b,ad}^{\mu\nu}\right) [A_{\mu},A^{\prime}_{\nu}]_{\star}
\Bigl|_{x^\prime=x}
\ee
where the terms ${\wt\Pi}_{a,ad}^{\mu\nu}$ and ${\wt\Pi}_{b,ad}^{\mu\nu}$
are now given by
\be
{\wt\Pi}_{a,ad}^{\mu\nu}={\rm tr}\int\frac{d^4p}{(2\pi)^4} S(p)\;{/\!\!\!b}
\gamma_5\;S(p)\gamma^{\mu}S(p)\gamma^{\nu}\;(\partial\times p)
\ee
and
\be
{\wt\Pi}_{b,ad}^{\mu\nu}={\rm tr}\int\frac{d^4p}{(2\pi)^4} S(p)\gamma^{\mu}
S(p)\;{/\!\!\!b}\gamma_5S(p)\gamma^{\nu}\;(\partial\times p)
\ee
We have
\be
{\wt\Pi}_{a,ad}^{\mu\nu}=b_{\lambda}\,{\rm tr}\int\frac{d^4p}{(2\pi)^4}
\frac{\partial\times p}{(p^2-m^2)^3}\bigl({/\!\!\!p}\gamma^\lambda\gamma_5
{/\!\!\!p}\gamma^\mu{/\!\!\!p}\gamma^\nu+m^2\,{/\!\!\!p}
\gamma^\lambda\gamma_5\gamma^\mu\gamma^\nu\bigr)
\ee
We use dimensional regularization in order to write
\be
\int\frac{d^{2w}p}{(2\pi)^{2w}}
\frac{p_\alpha p_\beta}{(p^2-m^2)^3}=
\frac{i (1-w)}{4(4\pi)^w}\frac{\Gamma(1-w)}{(m^2)^{2-w}}
g_{\alpha\beta}
\ee
and
\be
\int\frac{d^{2w}p}{(2\pi)^{2w}}
\frac{p_\alpha p_\beta p_\gamma p_\delta}{(p^2-m^2)^3}=
-\frac{i m^2}{8(4\pi)^w}\frac{\Gamma(1-w)}{(m^2)^{2-w}}
G_{\alpha\beta\gamma\delta}
\ee
We use the above results to get
\be\label{pia}
{\wt\Pi}_{a,ad}^{\mu\nu}=-\frac{m^2}{16\pi^2}
\left(\frac{4\pi\mu^2}{m^2}\right)^{2-w}\Gamma(1-w)(3-w)
\varepsilon^{\mu\nu\lambda\rho}b_\lambda{\bar\partial}_\rho
\ee
where ${\bar\partial}_\alpha=\theta_{\alpha\beta}\partial^\beta$.

The calculation involving ${\wt\Pi}_{b,ad}^{\mu\nu}$ is similar. It gives the
result ${\wt\Pi}_{b,ad}^{\mu\nu}=-{\wt\Pi}_{a,ad}^{\mu\nu}$, showing that
the contribution bilinear in the gauge field vanishes, as it did in the former
case.

In the noncommutative case there is another contribution, trilinear
in the gauge field, similar to Eq.~(\ref{i31}). It contributes with
\be
{\wt I}^{(1,3)}_{ad}[b,A]=\frac{i}{3}\int d^4x \sum_{n=1}^3
\Gamma_{n,ad}^{\,\mu\rho\nu}\;[[A_\mu,A^{\prime}_\rho]_{\star},
A^{\prime\prime}_\nu]_{\star}\Bigl|_{x^{\prime\prime}=x^{\prime}=x}
\ee
where
\be
\Gamma_{n,ad}^{\,\mu\rho\nu}=\Gamma_{n}^{\,\mu\rho\nu}(\partial\times p)
\ee
and $\Gamma_{n}^{\,\mu\rho\nu}$ stand for the three contributions given
by Eqs.~(\ref{G1}), (\ref{G2}), and (\ref{G3}). These results show that
there is no trilinear contribution independent of the derivative.

The above considerations lead to the result that there is no room for
Lorentz or CTP violation, despite the representation one chooses for the
gauge group. Our results show that there is no UV/IR mixing in the calculation
of the induced Chern-Simons term, even when non-planar diagrams are taken into
account, as it happens in the adjoint representation of the gauge group.

We summarize our work recalling that we have calculated the radiative
corrections induced by massive and charged Dirac fermions, interacting
with Abelian gauge field and including a non standard contribution
that violates Lorentz and CPT invariance. Our results show that
there is an intricate entanglement between infinitely large contributions
that come from integration in momentum space, and infinitely small
contributions that appear from the trace of Dirac matrices. These two
contributions compensate each other, and they do contribute to generate
a term that exactly cancels the finite term that appears from
the remaining contributions. Because of this intricate cancellation,
there is no room for radiative generation of the Chern-Simons-like
term. Thus, there is neither Lorentz nor CPT violation generated
radiatively. This result is valid under the general guidance
of dimensional regularization, despite the spacetime be commutative
or noncommutative.

We would like to thank CAPES, CNPq, PROCAD and PRONEX for partial support.

\end{document}